\documentclass[manuscript]{aastex}
\usepackage{graphicx}
\usepackage{subfigure}
\usepackage{dcolumn}
\usepackage{bm}
\usepackage{amssymb}

\begin{document}


\title{Influence of small scale $E_M$ and $H_M$ on the growth of large scale magnetic field}

\author{Kiwan Park}
\affil{Department of Astronomy and Physics, Ulsan National Institute of Science and Technology, South Korea 689798}

\begin{abstract}
We investigated the influence of small scale magnetic energy ($E_M$) and magnetic helicity ($H_M$) on the growth rate ($\gamma$) of large scale magnetic field ($\overline{{\bf B}}$). $H_M$ that plays a key role in MHD dynamo is a topological concept describing the structural properties of magnetic fields. So, it is not possible to differentiate the intrinsic properties of $H_M$ from the influence of $E_M$, and vice versa. However, to understand MHD dynamo the features of helical and nonhelical magnetic field should be made clear. For this, we made a detour: we gave each simulation set its own initial condition ($IC$, same $E_M$(0) and specific $H_M$(0) at $k_f=5$), and then drove the system with positive helical kinetic energy($k_f=5$). According to the simulation results, $E_M$(0), whether or not helical, increases the growth rate of $\overline{{\bf B}}$. The positive $H_M$(0) boosts the increased growth rate, but the negative $H_M$(0) decreases it. To explain these results two coupled equations of $H_M$ and $E_M$ were derived and solved using a simple approximate method. The equations imply that helical magnetic field generates the whole (helical and nonhelical) magnetic field but quenches itself. Nonhelical magnetic field also generates the whole magnetic field but quenches itself. The initially given $E_M$(0) modifies the electromotive force ($\langle {\bf v}{\bf \times} {\bf b}\rangle$, $EMF$) and generates new terms. The effects of these terms depend on the magnetic diffusivity $\eta$, position of initial conditions $k_f$, and time. But the influence disappears as time passes ($\sim e^{-\eta k_f^2 t}$), so the saturated magnetic fields are independent of the initial conditions.
\end{abstract}


\section{Introduction}
The evolutions of magnetic fields such as generation, amplification (dynamo), and annihilation (reconnection) are commonly observed in most celestial phenomena which include the interactions between magnetic fields and conducting fluids. The kinetic energy in the plasma motion can be transferred into the magnetic energy(dynamo), and this energy cascades toward smaller scale eddies and grows (small scale dynamo), or cascades toward larger scale ones and grows (large scale dynamo).  In MHD dynamo the role of helical kinetic motion (kinetic helicity, $\langle {\bf u} \cdot {\bf \omega}\rangle$, ${\bf \omega}={\bf \nabla} {\bf \times}{\bf u}$) is relatively clear: it generates the magnetic energy (helicity) and cascades the energy (helicity) to the larger scale magnetic eddies. However, the physical role and meaning of helical magnetic field (magnetic helicity, $H_M\equiv\langle {\bf A} \cdot {\bf B}\rangle$, ${\bf B}={\bf \nabla} {\bf \times}{\bf A}$) are not yet fully understood. $H_M$ is the topological measure of twist and linkage of magnetic field lines ($2\Phi_1\Phi_2$, $\Phi=\int_A {\bf B}\cdot d{\bf S}$, Krause \& R\"adler 1980, Moffatt 1978) in the minimum state of energy equilibrium. Helical magnetic field is called `force free field' because it makes Lorentz force (${\bf J} {\bf \times}{\bf B}$) $zero$ (Biskamp 2003). On the contrary, magnetic helicity is also related to the particle resonant scattering in the interplanetary magnetic fields when the handedness of helical magnetic field is the same as that of helical motion of a particle (Brown et al. 1999). And like magnetic energy($E_M$), $H_M$ is conserved in ideal plasmas. So the increasing large scale magnetic helicity leads to the generation and cascade of oppositely signed magnetic helicity toward smaller scale. Then, $H_M$ in the small scale (more exactly current helicity, $\langle {\bf j}\cdot {\bf b} \rangle=k^2H_M$) plays the role of constraining the growth of ${\overline{B}}$ fields (Blackman \& Field 2002). As an another example, quickly grown $H_M$ is ejected into the solar wind rather than being quenched; instead, the equal amount of oppositely signed $H_M$ is generated and stays in the sun. These last two examples and explanations are based on the conservation and redistribution of $H_M$, so that they show only partial features of $H_M$.\\


\noindent We are interested in the unique roles of helical \& nonhelical magnetic energy and their relation in MHD dynamo. But, it is not easy to answer to these questions because magnetic helicity assumes the existence of magnetic energy. In fact, $E_M$ can have arbitrary $|H_M|$ as long as realizability relation ($2E_M\geq |H_M|$, Frisch et al. 1975) is satisfied. So, we look for another indirect way to investigate $E_M$ and $H_M$ in MHD dynamo.\\

\noindent Before we go further, we need to make clear the statistical meaning of $H_M$. The correlation $\langle B_i(k)B_j(-k)\rangle$ can be represented by two invariants $E_M$ and $H_M$ like (Lesieur 1987, Park 2013, Yoshizawa 2011)
\begin{eqnarray}
\langle B_i({\bf k})B_j({\bf -k})\rangle = P_{ij}(k)\frac{E_M(k)}{4\pi k^2}+\frac{i\epsilon_{ijl}k_l}{8\pi k^2}H_M(k).\\
(P_{ij}(k)=\delta_{ij}-k_ik_j/k^2)\nonumber
\label{BiBj correlation}
\end{eqnarray}
In a homogeneous and isotropic (reflectionally symmetric) system, only the trace $\langle B_i({\bf k})B_i({\bf -k})\rangle$ ($\sim E_M$) survives. Off-diagonal term $\langle B_i({\bf k})B_j({\bf -k})\rangle$ ($i\neq j$) which is related to $H_M$ does not exist. This means that the second order correlation independent of translation and rotation (including reflection symmetry) can be described by the invariant variable $E_M$ (Robertson 1940). Actually most of the turbulence theories accept this assumption, and $E_M$ is used to describe the correlation $\langle B_i({\bf k})B_j({\bf -k})\rangle$. However, a system with such a strict condition is not common in nature. If there is a rotation, although the system is still isotropic, the reflection symmetry is broken so that $\langle B_i({\bf k})B_j({\bf -k})\rangle$ cannot be ignored. This off-diagonal term can be described by another invariant quantity, helicity. This formula implies helical fields are essentially related to the statistical correlations between `$B_i$' and `$B_j$' in an isotropic system. For example, current helicity $\langle {\bf J}\cdot {\bf B}\rangle$($=k^2H_M$) cannot exist without the off-diagonal correlation:
\begin{eqnarray}
\langle {\bf J}\cdot {\bf B}\rangle &=&
\int \langle {\bf J}({\bf k})\cdot {\bf B}(-{\bf k})\rangle\,d{\bf k} \nonumber \\
&=& \int \epsilon_{ijl}\, ik_j\langle B_l({\bf k}) B_i(-{\bf k})\rangle\, d{\bf k}\nonumber \\
&=&-\int \frac{1}{2}\epsilon_{ijl} \epsilon_{ilm}k_jk_m\,H_M(k)\,dk\nonumber \\
&=&\int k^2H_M(k)\,dk.
\label{proof of BiBj correlation}
\end{eqnarray}
But strictly speaking Eq.(1) is a description of the second correlation tensor rather than a conservation law. Although $E_M$ is described as a trace in this formula, it can include helical magnetic energy ($kH_M/2$) and nonhelical magnetic energy ($E_M-kH_M/2$).

\section{Problem to be solved and methods}
The main aim of this paper is to figure out the effect of initial conditions($H_M$(0) \& $E_M$(0)) in small scale on the growth of large scale MHD dynamo. Pouquet et al. (1976) derived the equations of $E_M$, $H_M$, $E_V$, and $H_V$ using EDQNM. The results show the features of the variables and explain how the inverse cascade of $E_M$ and $H_M$ with $\alpha$ coefficient occurs. But the physical difference between $E_M$ and $H_M$ in MHD dynamo is not clearly shown. Driving the system with the mixed helical and nonhelical kinetic energy, Maron \& Blackman 2002 tried to see the effects of various helicity ratio. The results show the mixed effect of partially helical and nonhelical kinetic energy, but the influence of $H_M$ or $E_M$ on MHD dynamo is not shown. In Ref. Park (2013) and Park et al. (2013), it was shown that $H_M$(0) and $E_M$(0) in the large scale boosted the generation of ${\overline{B}}$ field. But the work was chiefly focused on the influence of $E_M(0)$. So, we need more detailed analytic and experimental work which can show the effect of $H_M$ and $E_M$. For this purpose we prepared for some simulation sets. Magnetic energy $E_M$ with a fractional helicity($fhm$) drove a system as a precursor simulation ($k_f$=5, t$<$0.005, one simulation step) to generate $E_M$(0) and $H_M$(0) in the system. Then fully helical kinetic energy ($fhk$=1.0) was injected into the kinetic eddy at $k_f$=5 (helical kinetic forcing $HKF$) to drive the system as a main simulation.\\

\noindent All simulations were done with high order finite difference Pencil Code(Brandenburg 2001) and the message passing interface(MPI) in a periodic box of spatial volume $(2 \pi)^3$ with mesh size $256^3$. The basic equations solved in the code are,
\begin{eqnarray}
\frac{D \rho}{Dt}&=&-\rho {\bf \nabla} \cdot {\bf u}\\
\frac{D {\bf u}}{Dt}&=&-c_s^2{\bf \nabla} \mathrm{ln}\, \rho + \frac{{\bf J}{\bf \times} {\bf B}}{\rho}+\nu\big({\bf \nabla}^2 {\bf u}+\frac{1}{3}{\bf \nabla} {\bf \nabla} \cdot {\bf u}\big)+{\bf f}\\
\frac{\partial {\bf A}}{\partial t}&=&{\bf u}{\bf \times} {\bf B} -\eta\,{\bf \nabla}{\bf \times}{\bf B}.
\label{MHD equations in the code}
\end{eqnarray}
$\rho$: density; $\bf u$: velocity; $\bf B$: magnetic field; $\bf A$: vector potential; ${\bf J}$: current density;  $D/Dt(=\partial / \partial t + {\bf u} \cdot {\bf \nabla}$): advective derivative; $\eta$: magnetic diffusivity(=$c^2/4\pi \sigma$, $\sigma$: conductivity); $\nu$: kinematic viscosity(=$\mu/\rho$, $\mu$: viscosity); $c_s$: sound speed. Velocity is expressed in units of $c_s$, and magnetic fields in units of $(\rho_0\,\mu_0)^{1/2}c_s$($B=\sqrt{\rho_0\,\mu_0}v$). $\mu_0$ is magnetic permeability and $\rho_0$ is the initial density. Note that $\rho_0\sim \rho$ in the weakly compressible simulations. These constants $c_s$, $\mu_0$, and $\rho_0$ are set to be `1'. In the simulations $\eta$ and $\nu$ are 0.006. To force the magnetic eddy($HMF$), forcing function `${\bf f}(x,t)$' is placed at Eq.(5) first; and then `${\bf f}$' is placed at Eq.(4) to drive the momentum equation($HKF$). ${\bf f}(x,t)$ is represented by $N\,{\bf f}_0(t)\, exp\,[i\,{\bf k}_f(t)\cdot {\bf x}+i\phi(t)]$($N$: normalization factor, ${\bf f}_0$: forcing magnitude, ${\bf k}_f(t)$: forcing wave number). The amplitude of magnetic forcing function($f_0$) was $0.01$ with various magnetic helicity ratios modifying $fhm$ during $HMF$; and $f_0$ of $HKF$ was $0.07$. The variables in pencil code are independent of a unit system. For example, if the length of cube box $L$ is $2\pi$ and $u_{rms}$ is $\sim 0.2$ after $t=3$, these can be interpreted as $L$ = $2\pi$ m, $u_{rms}\sim 0.2$ m/s, $t$ = 3 s, or $L$ = $2\pi$ pc, $u_{rms}\sim 0.2$ pc/Myr, $t$ = 3 Myr. And for the theoretical analysis, we use semi analytic and statistical methods. The equations of $E_M$ \& $H_M$ with the solutions are derived again using an approximation like FOSA (first order smoothing approximation, Moffatt 1978).

\begin{figure}
   \center
   {
     \includegraphics[width=8.0 cm]{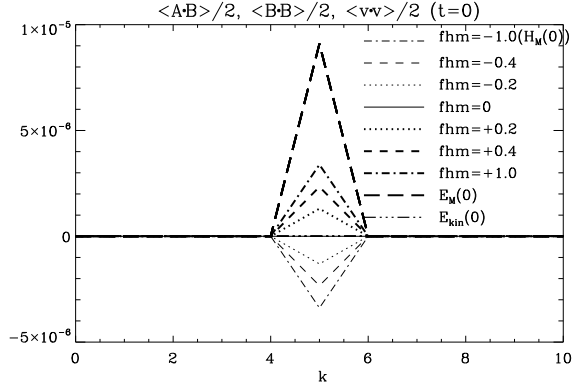}
     \label{1a}
}
\caption{$H_M(0)/2$, $E_M(0)$, and $E_{kin}(0)$. Thick long-dashed line of the highest peak is the common $E_M$(0) for all simulation sets. The other lower peak lines indicate $H_M(0)/2$ for each separate simulation. The horizontal line passing through `0' is $E_{kin}$(0) for all simulations.}

\end{figure}

\begin{figure*}
\center
\mbox{%
   \subfigure[]
   {
     \includegraphics[width=8cm]{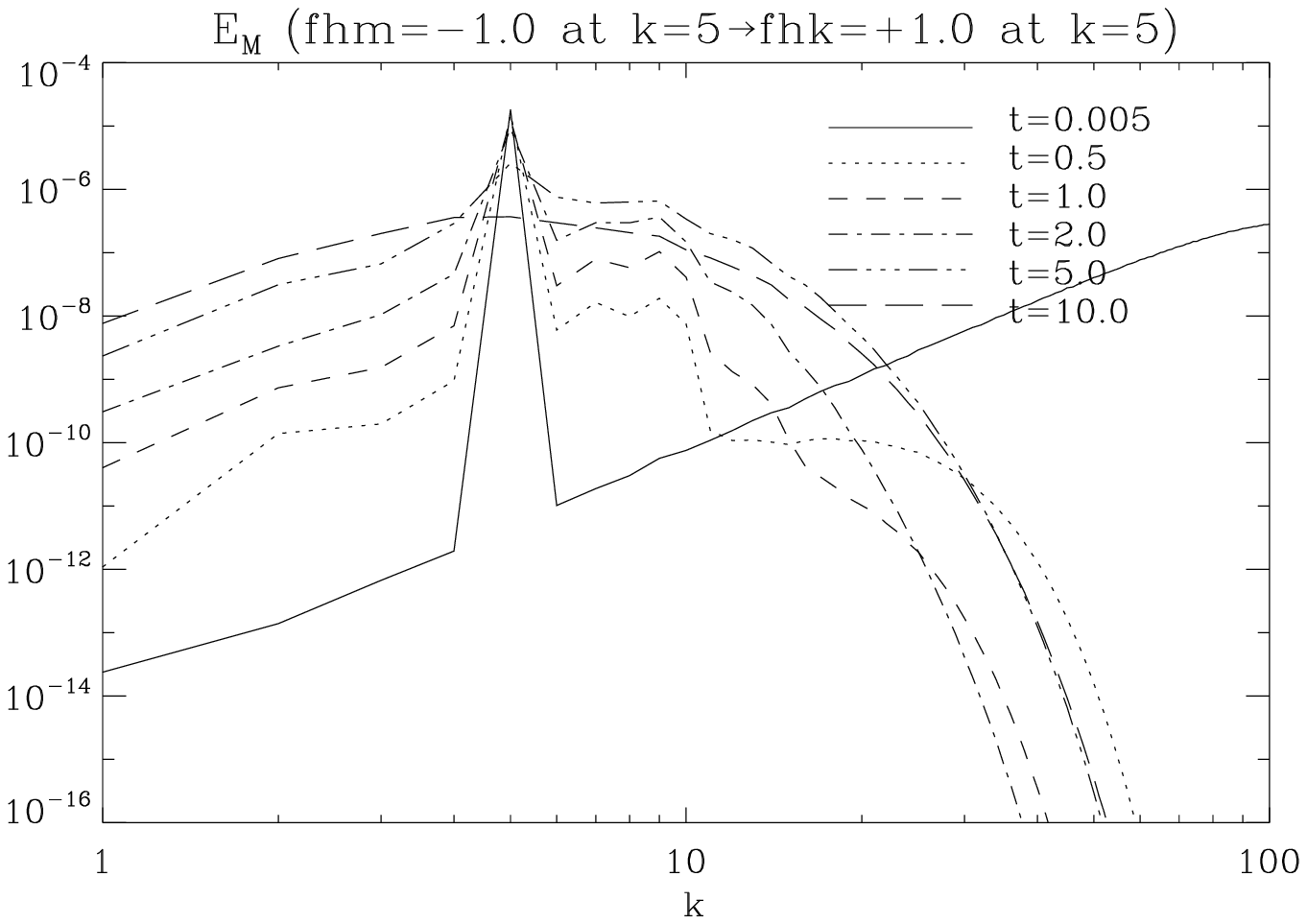}
     \label{2anew}
               }\,
   \subfigure[]
   {
     \includegraphics[width=8cm]{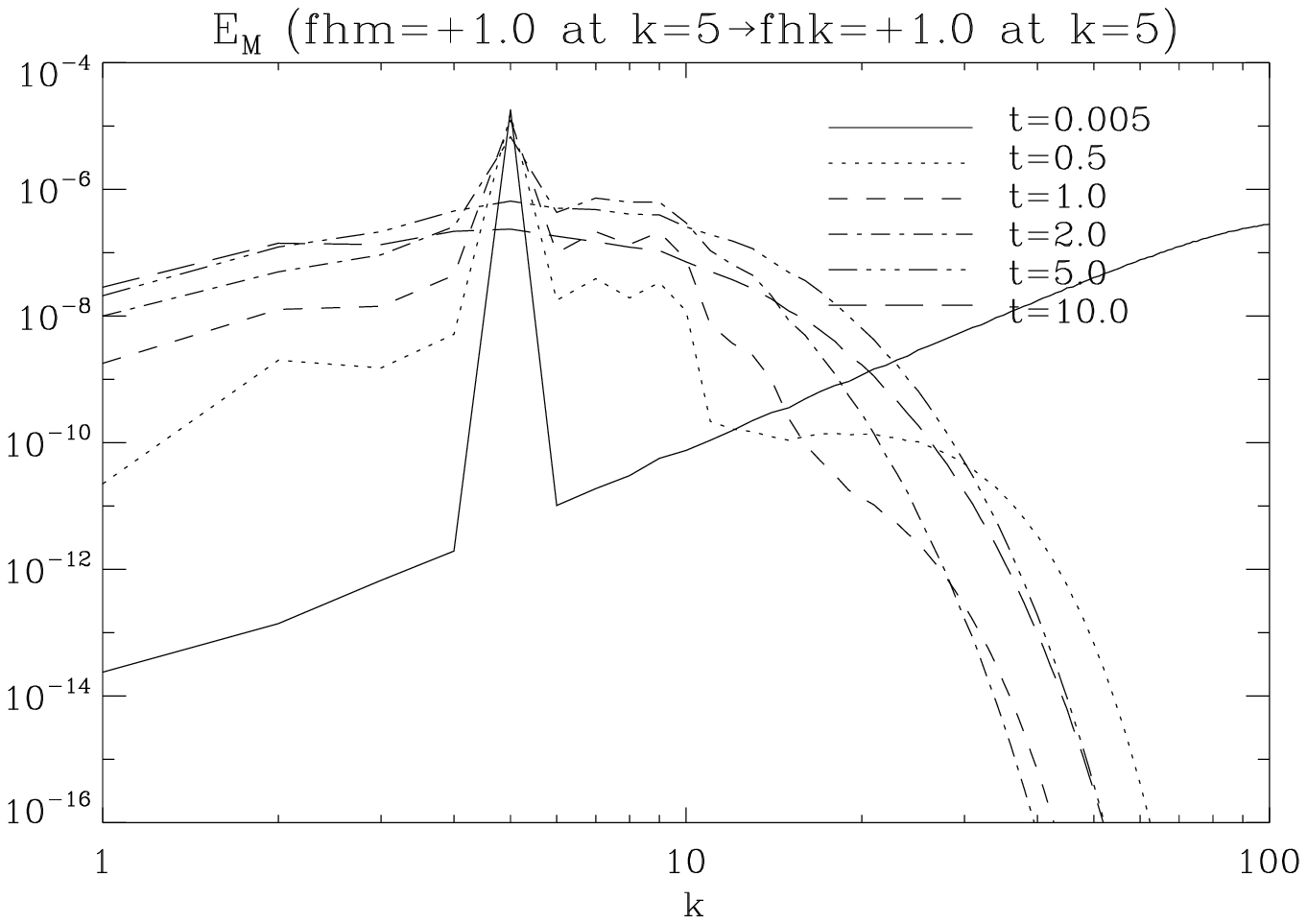}
     \label{2bnew}
               }
}

\caption{(a) $E_M$ spectrum with the negative initial $H_M$(0) (b) $E_M$ with the positive initial $H_M$(0)}

\end{figure*}

\begin{figure*}
\center
\mbox{%
   \subfigure[]
   {
     \includegraphics[width=7.5cm]{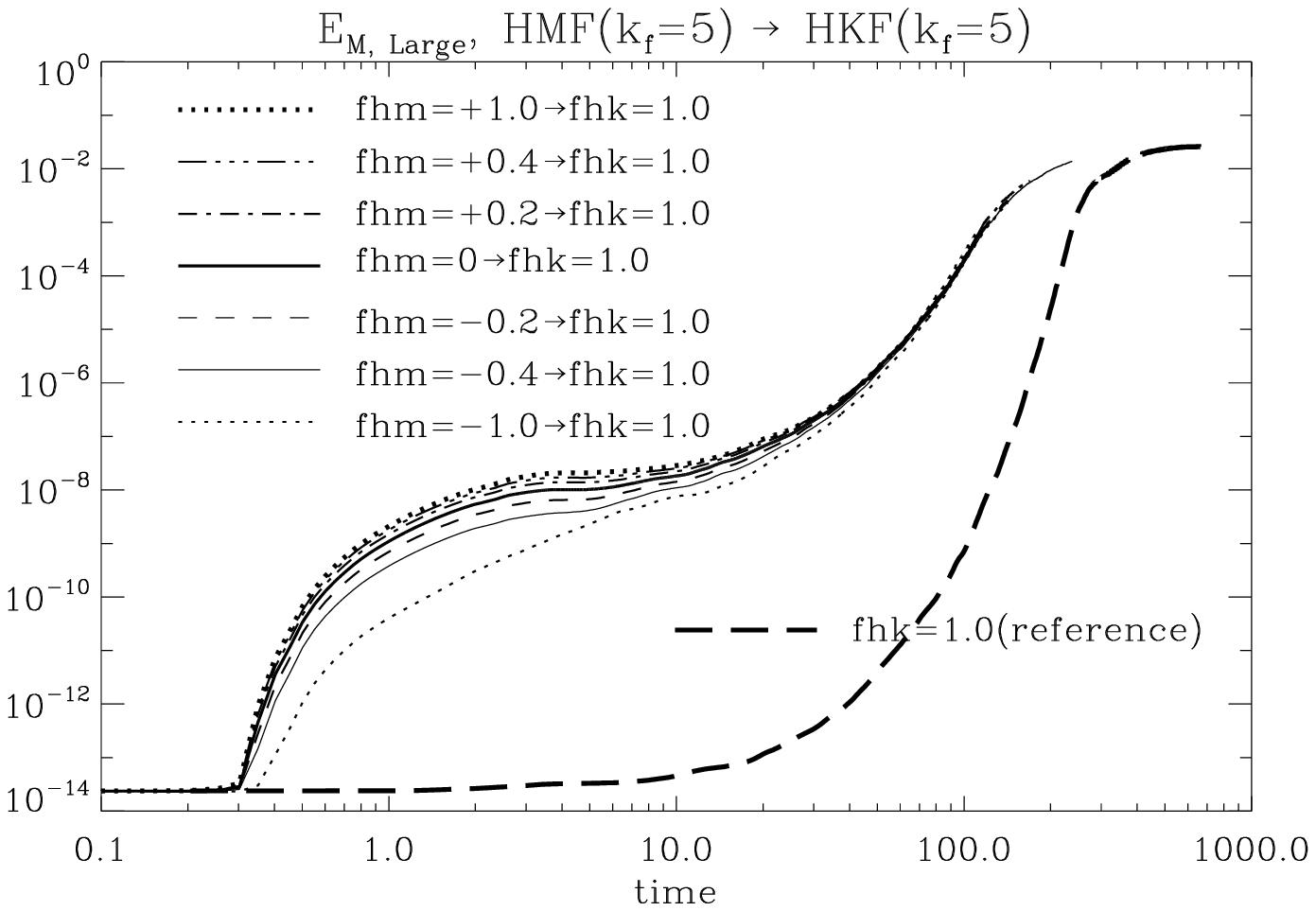}
     \label{2a}
               }\,
   \subfigure[]
   {
     \includegraphics[width=7.5cm]{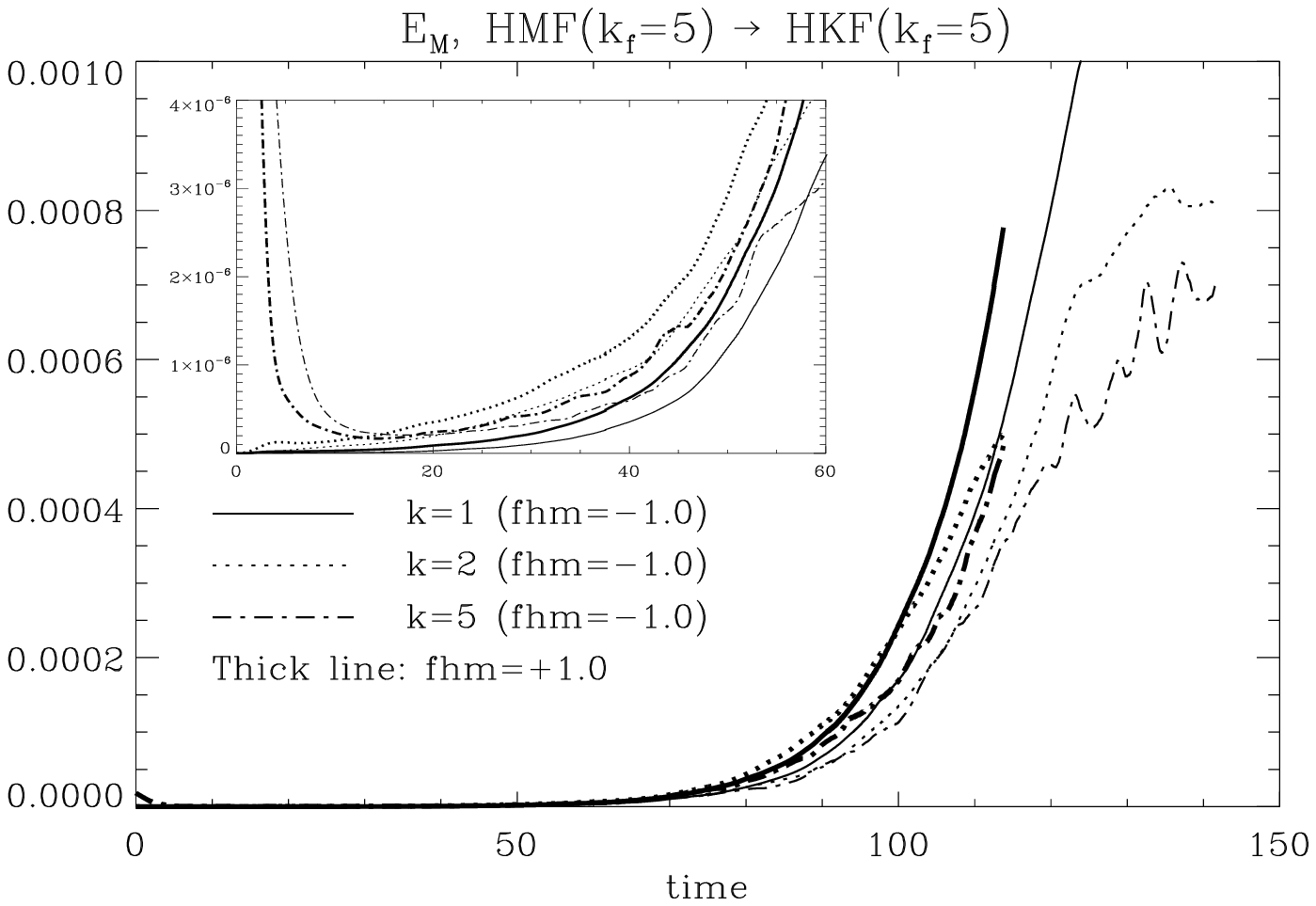}
     \label{2b}
               }
}

\mbox{
   \subfigure[]
   {
     \includegraphics[width=7.5cm]{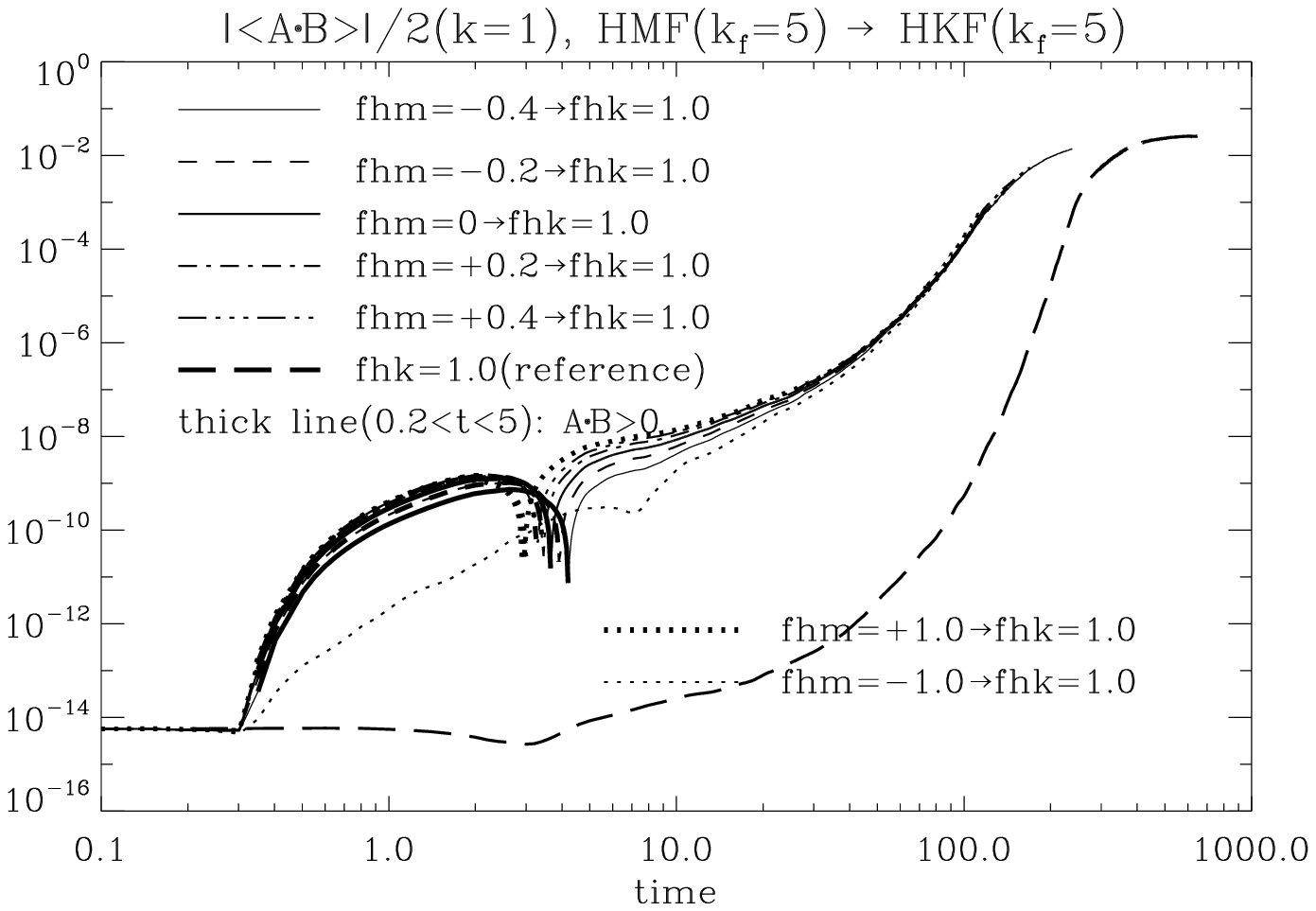}
     \label{2c}
               }\,
   \subfigure[]
   {
     \includegraphics[width=7.5cm]{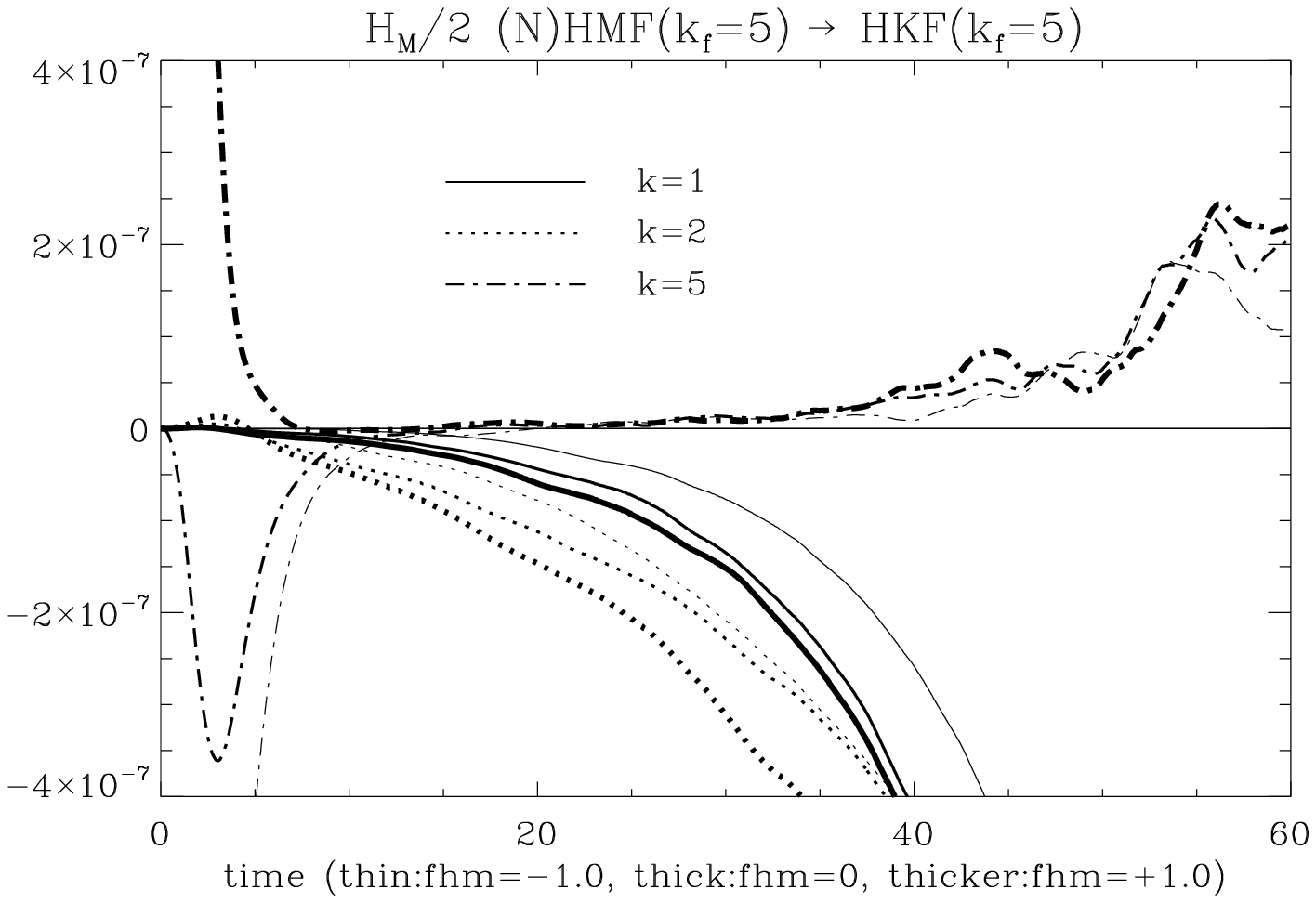}
     \label{2d}
               }

}

\mbox{
   \subfigure[]
   {
     \includegraphics[width=7.5cm]{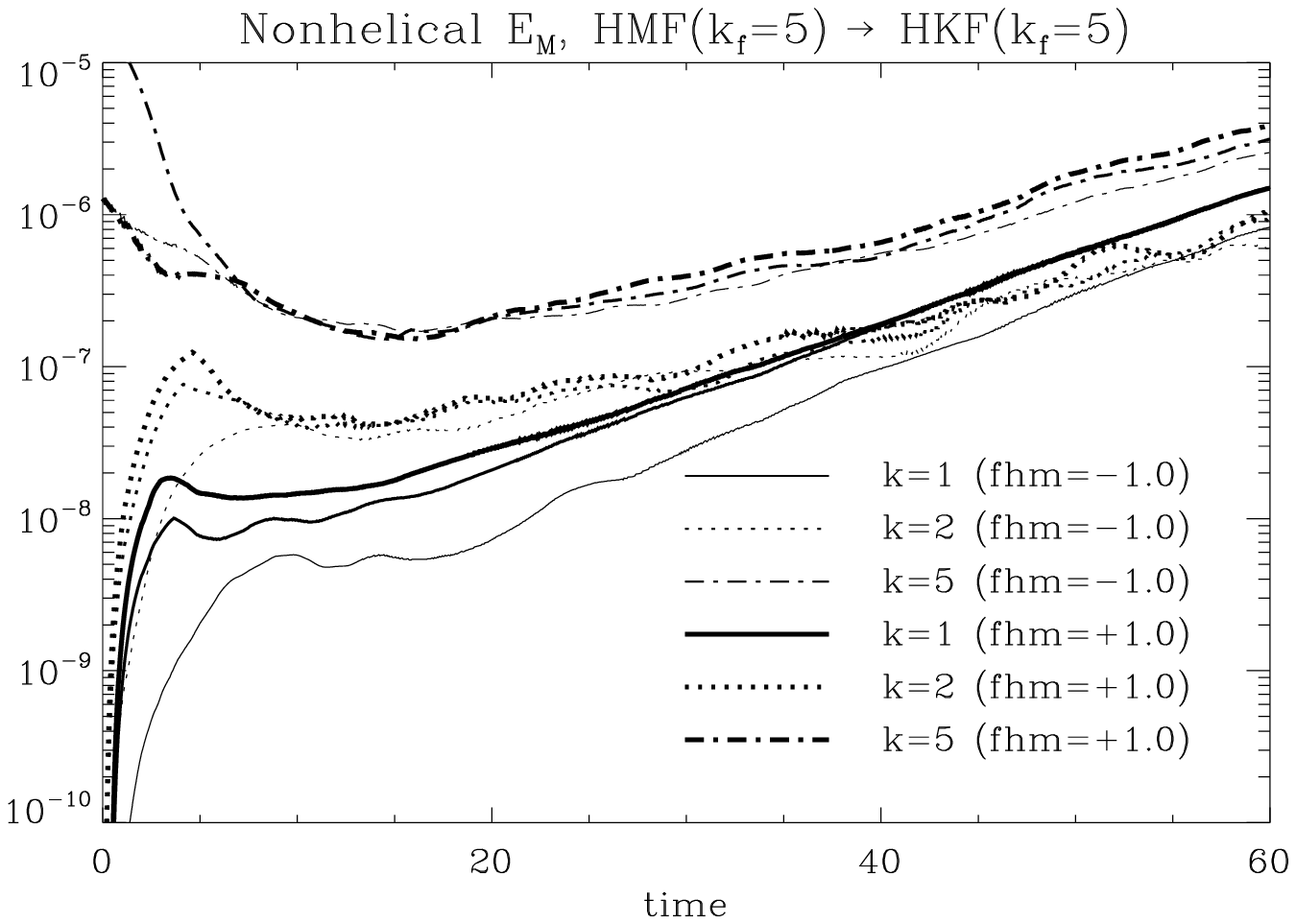}
     \label{2e}
               }\,
   \subfigure[]
   {
     \includegraphics[width=7.5cm]{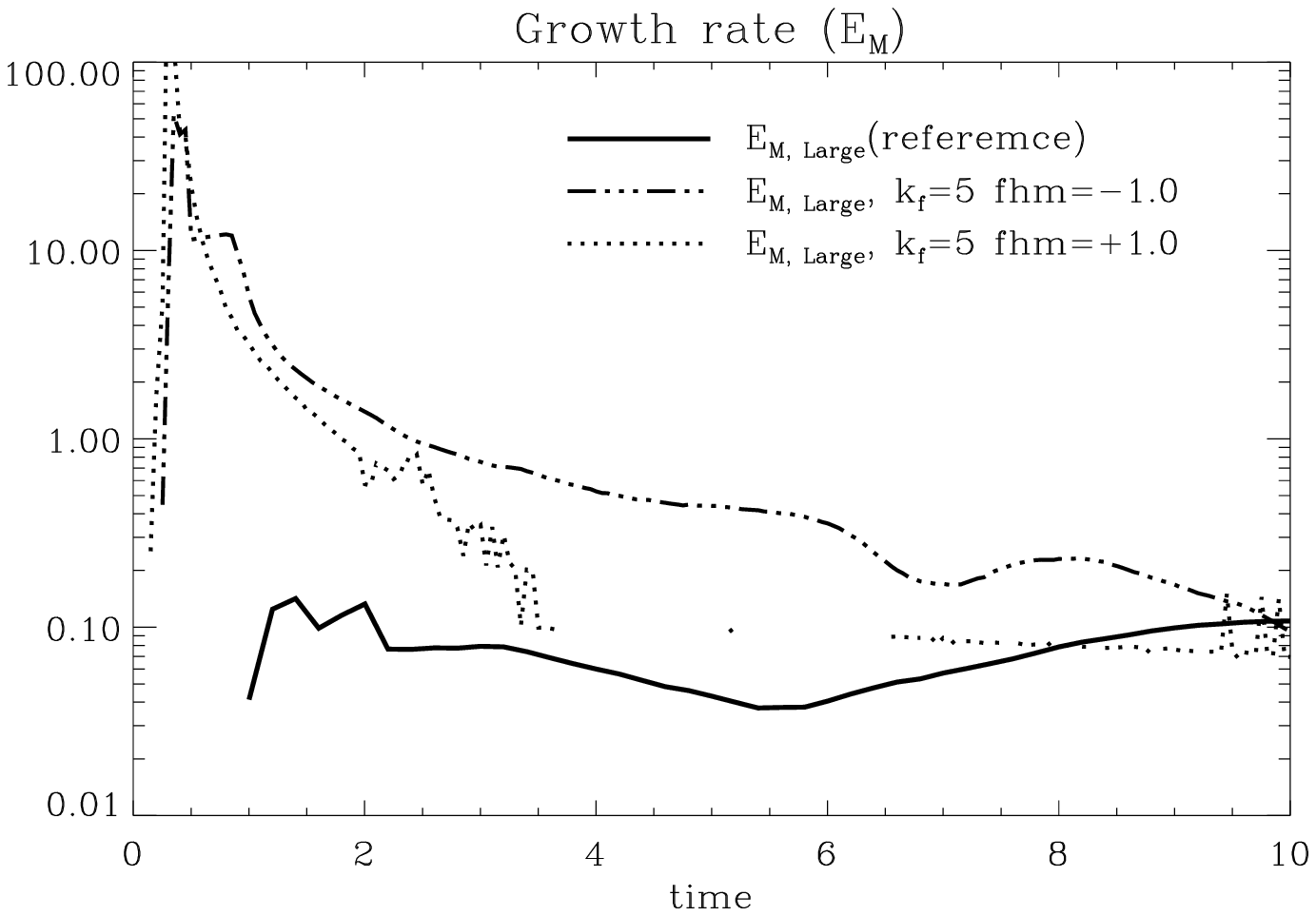}
     \label{2f}
               }

}

\caption{(a) The plots of $E_{M,L}$ with various $H_M$(0) at $k_f$=5. (b) $E_M$ at $k$=1, 2, and 5. Thin lines indicate $E_M$ with negative $H_M$(0)($fhm=-1$), thick lines are for $E_M$ with zero $H_M$(0)($fhm=0$), and thicker lines are for $E_M$ with positive $H_M$(0)($fhm=+1$). The small box includes the magnified plots of $E_M$. (c) $|H_{M,L}|$/2 with various $H_M$(0). $H_{M,L}$ is negative when the system is driven by the positive $\langle {\bf v}\cdot {\bf \omega}\rangle$, but it is positive (thick lines) at $0.3<t<8$. (d) $H_M$ of $k$=1, 2, and 5. (e) Nonhelical $E_M$($E_M$-k$|H_M|$/2). Thin line is for fhm=-1, thick line is for fhm=0, and thicker line is for fhm=1 (f) Growth ratio $\gamma$ ($d\,\mathrm{log}\, E_M/dt)$}

\end{figure*}

\begin{figure*}
\center
\mbox{%
   \subfigure[]
   {
     \includegraphics[width=9.5cm]{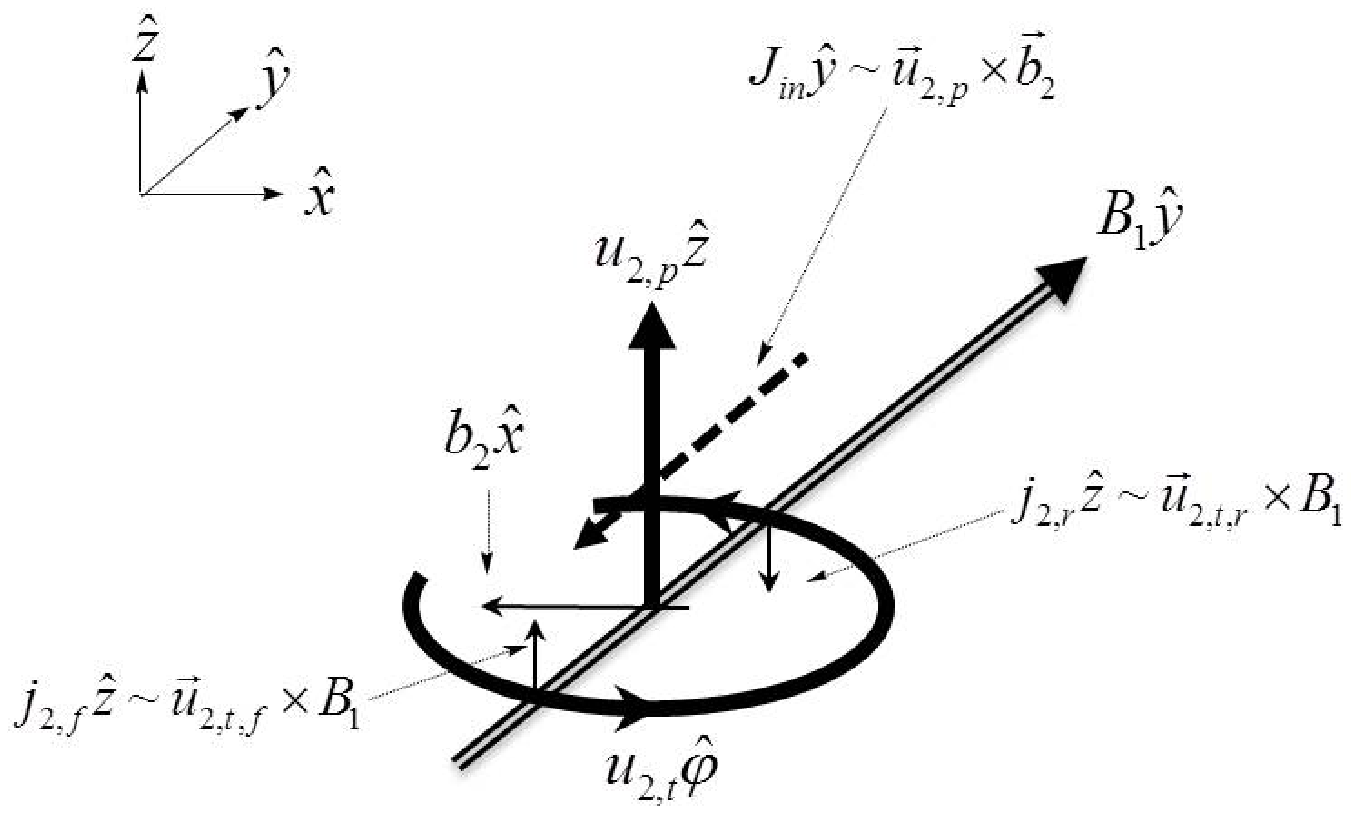}
     \label{4a}
               }\,
   \subfigure[]
   {
     \includegraphics[width=5.5cm]{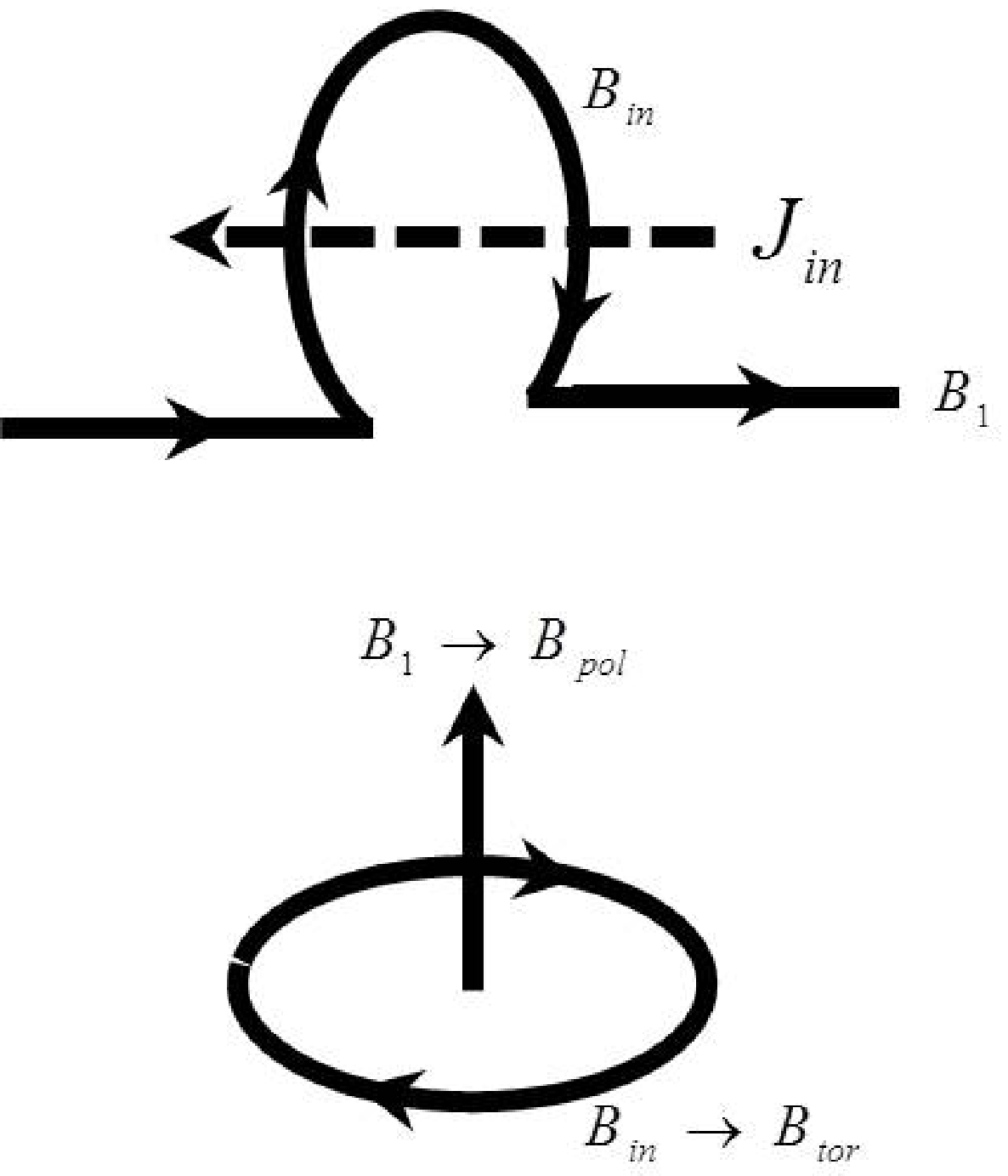}
     \label{4b}
               }

}
\caption{The generation of ${\bf J}_{in}$ is in consequence of the interaction between ${\bf u}$ and ${\bf B}_1$. ${\bf J}_{in}$ produces ${\bf B}_{in}$ from ${\bf B}_1$, both of which develop a left handed magnetic field structure.}
\end{figure*}

\section{Simulation results}
\noindent Fig.1 shows the initial distributions of $H_M$(0), $E_M$(0), and $E_{kin}(0)$. This figure includes the eight simulation sets of $fhm$=$\pm$1.0, $fhm$=$\pm$0.4, $fhm$=$\pm$0.2, and $fhm$=0. $H_M$(0)/2 at $k$=5 are $\pm3.38{\bf \times}10^{-6}$($fhm$=$\pm1$), $\pm2.34{\bf \times}10^{6}$($fhm$=$\pm0.4$), $\pm1.31{\bf \times}10^{-6}$($fhm$=$\pm0.2$), $-1.35{\bf \times}10^{-10}$($fhm$=0). However, $E_M(0)$ of each case is consistently the same ($1.82{\bf \times} 10^{-5}$). $E_M$(0) and $H_M$(0)/2 of the reference simulation are actually `$zero$' ($5.36{\bf \times} 10^{-12}$ and $-4.68{\bf \times} 10^{-14}$). $E_{kin}$ is not influenced by the preliminary magnetic forcing so that all simulation sets initially have `$zero$' $E_{kin}$(0).\\

\noindent Fig.\ref{2anew} is the evolving $E_M$ spectrum which has the negative $H_M$(0) ($fhm=-1$, $k$=5, $t=0$). Fig.\ref{2bnew} has the same conditions except the positive $H_M$(0) ($fhm=+1$) at $k$=5. Here, the peak of $E_M$ drops faster; but, the growth rate of $E_M$ ($k<5$) is larger. Negative $H_M$ which is generated by the positive $\langle {\bf v}\cdot \omega \rangle$ is injected into the positive $H_M$(0) at $k$=5.\\

\noindent Fig.\ref{2a} shows the growth rate of $E_{M,L}$ increases in proportion to $H_M(0)$. However, the comparison of plots indicates $E_M$(0) is a more important factor in the growth rate. $E_M(0)$ and $H_M(0)$ of the reference $HKF$ are $5.02{\bf \times}10^{-12}$ and $-9.02{\bf \times}10^{-14}$, and those of negative $H_M(0)$ ($fhm=-1$) are $1.82{\bf \times}10^{-5}$ and $-6.77{\bf \times}10^{-6}$. In spite of much smaller $H_M(0)$, the simulation with $fhm=-1$ has even larger growth rate than that of the reference. Moreover, the large scale magnetic field is saturated faster. But, the saturated values are the same.\\

\noindent Fig.\ref{2b} includes the detailed evolutions of $E_M$ at $k$=1, 2, 5 for $fhm=0, \,-1.0,\,+1.0$. The plot shows $E_M$ with the positive $H_M$(0) at $k$=5 decreases faster than that of the negative $H_M$(0) when the negative $H_M$ is injected into the system. This fast drop of $E_{M}$ at $k$=5 leads to the larger growth of $E_{M}$ at $k$=1, 2.\\

\noindent Fig.\ref{2c} is to compare the growth rate of $|H_{M,L}|$ with various initial magnetic helicity. The growth rate is also proportional to $E_M$(0) \& $H_M$(0). Thick lines($0.3<t<8$) indicate $H_{M,L}$ is positive in this time regime, and thin lines are negative $H_{M,L}$. The cusps are points where the positive $H_M$ turns into a negative one. This positive $H_M$ is thought to be caused by the tendency of conserving $H_{M, tot}$ against the injected negative $H_M$. However, if the magnitude of negative $H_M$(0) is large($fhm=-1$) enough or $E_M$ is not so large(reference $HKF$), $H_{M,L}$ does not change its sign.\\

\noindent Fig.\ref{2d} shows the evolving profiles of $H_M$ at $k$=1, 2, 5. $H_M$ at $k$=1, 2 are negative. But $H_M$ of $k$=5 turns into a positive value regardless of the sign of $H_M$(0), which is due to the back reaction of the larger scale magnetic field. While $H_M$(0) at $k$=5 is positive (but $H_M$ at $k$=1, 2 is negative), $H_M$(0) decreases faster than the negative $H_M$(0). And this fast decrease of $H_M$ boosts the growth of $|H_M|$ at $k$=1, 2. In contrast, for the negative $H_M$(0) at $k$=5, the injected (negative) $H_M$ mitigates the decreasing speed of $|H_M|$ at $k$=5 and growing speed of $|H_M|$ at $k$=1, 2. Similarly for $fhm=0$ $H_M$ first drops. However, as the magnitude of $\overline{B}$ field grows, the diffusion of positive $H_M$ from large scale makes $H_M$ at $k$=5 grow to be positive.\\

\noindent Fig.\ref{2e} includes the evolving profiles of nonhelical $E_M$ ($E_M-k|H_M|/2$). Larger $H_M(0)$ at $k$=5 leads to the larger growth ratio of nonhelical $E_M$ at $k$=1, 2. And when the diffusion of energy from larger scale grows, the flat profile (in nonhelical $E_M$) at $k$=5 shows up ($\sim 2<t<\sim 7$). And then the profiles of $k$=5 for each case evolve together independent of the different evolutions of large scale ${\bf B}$ fields for a while($t<\sim 20$). The profiles of $E_M$ at $k$=2 also show the similar, but short pattern.\\

\noindent Fig.\ref{2f} includes the growth ratios of large scale $B$ field for $fhm=+1.0$, $fhm=-1.0$, and the reference simulation. The positive $H_M(0)$ causes the highest $\gamma$ in the early time regime. Also, the comparison of growth ratio between fhm=-1 and reference HKF implies that $E_M(0)$ is a more important factor than $H_M(0)$ in MHD dynamo.

%
%

\section{Analytic solutions to $H_M$ and $E_M$}
\noindent For the analytic approach, we use more simplified equations than Eq.(3)-(5). If we combine Faraday's law $\partial {\bf B}/\partial t=-c{\bf \nabla}{\bf \times}{\bf E}$ and Ohm's law ${\bf J}=\sigma({\bf E}+1/c{\bf U}{\bf \times} {\bf B}$), we get the magnetic induction equation:
\begin{eqnarray}
\frac{\partial {\bf B}}{\partial t}={\bf \nabla} {\bf \times} ({\bf U}{\bf \times} {\bf B}) + \eta {\bf \nabla}^2{\bf B}.
\label{magnetic induction 1}
\end{eqnarray}
All variables can be split into the mean and fluctuating values like ${\bf U}={\bf \overline{U}}+{\bf u}\rightarrow {\bf u}$(${\bf \overline{U}}\equiv$0, Galilean transformation) and ${\bf B}={\bf \overline{B}}+{\bf b}$. Then, the magnetic induction equation for ${\bf \overline{B}}$ field becomes (Krause \& R\"adler 1980),
\begin{eqnarray}
\frac{\partial {\bf \overline{B}}}{\partial t}&=&{\bf \nabla} {\bf \times} \langle {\bf u}{\bf \times} {\bf b}\rangle + \eta {\bf \nabla}^2{\bf \overline{B}}\\
&\sim&{\bf \nabla} {\bf \times} \alpha {\bf \overline{B}} + (\eta+\beta) {\bf \nabla}^2{\bf \overline{B}}.
\label{magnetic induction 2}
\end{eqnarray}
(Here, the electromotive force $EMF$ $\langle {\bf u}{\bf \times} {\bf b}\rangle$ was replaced by $\alpha{\bf \overline{B}}-\beta {\bf \nabla} {\bf \times} {\bf \overline{B}}$. $\alpha=1/3\int^t (\langle {\bf j}\cdot {\bf b}\rangle-\langle {\bf u}\cdot {\bf \omega}\rangle)d\tau, \footnote{The helicity terms in `$\alpha$' indicate the MHD system is isotropic without the reflection symmetry.}\,\, \beta=1/3\int ^t\langle u^2\rangle\,d\tau$, Moffatt 1978)\\

\noindent $E_M$(t) or $H_M$(t) can be derived using EDQNM(Pouquet et al. 1976), but the same equations can be derived using a mean field method (Park \& Blackman 2012(913P), Park \& Blackman 2012(2120P)). With Eq.(\ref{magnetic induction 2}), we get $\partial H_M / \partial t$(Blackman \& Field 2002, Krause \& R\"adler 1980):
\begin{eqnarray}
\frac{\partial}{\partial t}\langle{\bf \overline{A}}\cdot {\bf \overline{B}}\rangle&=&
2\langle{\bf \overline{\xi}}\cdot {\bf \overline{B}}\rangle-2\eta \langle{\bf \overline{B}}\cdot{\bf \nabla} {\bf \times} {\bf \overline{B}}\rangle\nonumber\\
&=&2\alpha\langle {\bf \overline{B}}\cdot {\bf \overline{B}} \rangle-2(\beta+\eta)\langle {\bf \overline{B}}\cdot{\bf \nabla}{\bf \times}{\bf \overline{B}}\rangle
\label{Hm1}
\end{eqnarray}
Considering helicity is a pseudoscalar, this equation can be represented like
\begin{eqnarray}
\frac{\partial}{\partial t}H_{M,L}=4\alpha E_{M,L}-2k^2(\beta+\eta)H_{M,L}\,\,(k=1).
\label{Hm2}
\end{eqnarray}
Also $\partial_t E_{M,L}$ can be derived from Eq.(\ref{magnetic induction 2}).
\begin{eqnarray}
\frac{\partial }{\partial t}\frac{1}{2}\langle \overline{B}^2\rangle&=&\langle{\bf \overline{B}}\cdot{\bf \nabla}{\bf \times} {\bf\xi}\rangle-\frac{c}{\sigma}\langle{\bf \overline{B}}\cdot {\bf \nabla}{\bf \times}{\bf \overline{J}}\rangle\nonumber\\
&=&\langle\alpha {\bf \overline{B}}\cdot {\bf \nabla} {\bf \times}{\bf \overline{B}}\rangle-\langle\beta {\bf \nabla} {\bf \times} {\bf \overline{B}}\cdot {\bf \nabla} {\bf \times}{\bf \overline{B}}\rangle-\frac{c}{\sigma}\langle{\bf \overline{J}}\cdot {\bf \nabla}{\bf \times}{\bf \overline{B}}\rangle.
\label{Em1}
\end{eqnarray}
In Fourier space,
\begin{eqnarray}
\frac{\partial }{\partial t}E_{M,L}
&=&\alpha k^2\langle{\bf \overline{A}}\cdot {\bf \overline{B}}\rangle-k^2\big(\beta+\eta\big)\langle{\bf \overline{B}}^2\rangle\nonumber\\
&=&\alpha k^2H_{M,L}-2k^2\big(\beta+\eta\big)E_{M,L}.\,\,(k=1)
\label{Em2}
\end{eqnarray}
$\overline{{\bf B}}$ or $E_{M,L}$ itself includes the helical and nonhelical part, but the nonhelical one is excluded in $\langle \overline{{\bf A}}\cdot \overline{{\bf B}}\rangle$ or $\langle \overline{{\bf B}}\cdot {\bf \nabla \times} \overline{{\bf B}}\rangle$.\\

\noindent Helical magnetic field in small scale constrains the growth of $\overline{{\bf B}}$ field, and nonhelical magnetic field ($\sim E_M-kH_M/2$) restricts the plasma motion through Lorentz force $\langle {\bf J} \times {\bf B}\rangle$(=${\bf B}\cdot {\bf \nabla B}-{\bf \nabla}B^2/2$). Eq.(\ref{Hm1}) and Eq.(\ref{Em1}) show additional relations between $H_{M,L}$ and $E_{M,L}$: the growing correlation $\langle B_iB_j \rangle$ leads to the increase of $E_M$, and growing $E_M$ increases the correlation $\langle B_iB_j \rangle$, but at the same time the dissipation effect of $E_M$ ($H_M$) grows with increasing $E_M$ ($H_M$). Besides, magnetic energy in small scale affects the electromotive force $\langle {\bf v} \times {\bf b} \rangle$ to change the growth ratio of $\overline{{\bf B}}$ field whether the field is helical or not.\\

\noindent $\partial E_M/\partial t$ and $ \partial H_M/\partial t$
have two normal mode solutions $\langle {\bf A}\cdot {\bf B}\rangle+\langle {\bf B}\cdot {\bf B} \rangle$ and $\langle {\bf A} \cdot {\bf B}\rangle-\langle {\bf B}\cdot {\bf B}\rangle$. Then two exact solutions are (Park 2013),
\begin{eqnarray}
2H_{M,L}(t)&=&(H_{M,L0}+2E_{M,L0})e^{2\int^t_0(\alpha-\beta-\eta)d\tau}\nonumber\\
&+&(H_{M,L0}-2E_{M,L0})e^{-2\int^t_0(\alpha+\beta+\eta)d\tau},
\label{EmHmSolution1}
\end{eqnarray}
\begin{eqnarray}
4E_{M,L}(t)&=&(H_{M,L0}+2E_{M,L0})e^{2\int^t_0(\alpha-\beta-\eta)d\tau}\nonumber\\
&-&(H_{M,L0}-2E_{M,L0})e^{-2\int^t_0(\alpha+\beta+\eta)d\tau}.
\label{EmHmSolution2}
\end{eqnarray}
$E_{M,L}(t)$ and $H_{M,L}(t)$ proportionally depend on $E_{M,L}(0)$ and $H_{M,L}(0)$, and their evolutions also depend on $\int_0^t (\alpha-\beta-\eta)\, d\tau$ and $\int_0^t (\alpha+\beta+\eta)\, d\tau$. The effect of initial small scale fields shows up while large scale ${\bf B}$ field is weak. The profile of small scale eddies becomes subordinate to the large scale magnetic field in a few eddy turnover times. While $\beta$ and $\eta$ are always positive, $\alpha$ is negative when the system is driven by the positive helical velocity field. Thus, the second terms on the right hand side in Eq.(\ref{EmHmSolution1}) and Eq.(\ref{EmHmSolution2}) dominantly decide the profiles of $E_{M,L}$ and $H_{M,L}$. And negative `$H_{M,L0}-2E_{M,L0}$' indicates that the evolving $E_M$ is positive but $H_{M,L}$ is negative.\\

\noindent The initially given $E_M$(0)($\sim b_i^2(0)$ at $k_f=5$) changes $EMF$. Since the interaction between ${\bf u}$ and ${\bf b}_i(0)$ can be ignored in the very early time regime, the magnetic induction equation is
\begin{eqnarray}
\frac{\partial {\bf b}_i}{\partial t}
\approx\eta{\bf \nabla}^2{\bf b}_i.
\label{MagIndSmall1}
\end{eqnarray}
In Fourier space,
\begin{eqnarray}
\frac{\partial {\bf b}_i}{\partial t}
\approx-\eta k_f^2{\bf b}_i\Rightarrow {\bf b}_i(t)={\bf b}_i(0)\,e^{-\eta k^2_ft}.
\label{MagIndSmall2}
\end{eqnarray}
Total magnetic field is composed of ${\bf \overline{B}}(k=1)$, ${\bf b}_i(k=5)$, and ${\bf b}(2\leq k \leq k_{max})$. Strictly speaking ${\bf b}_i$ is in the small scale. However, since such large $E_M$(0) decreases quickly before ${\bf u}$ grows enough to interact with ${\bf b}_i$, we can think ${\bf b}_i$ evolves independently (Fig.2).

\noindent Ignoring dissipation term for simplicity, the approximate small scale magnetic field ${\bf b}$ is
\begin{eqnarray}
\frac{\partial {\bf b}}{\partial t}
&=&{\bf \nabla}{\bf \times}({\bf u}{\bf \times}\overline{\bf B})+{\bf \nabla}{\bf \times}({\bf u}{\bf \times}{\bf b}_i).
\label{MagIndSmall3}
\end{eqnarray}
This equation indicate $EMF$ ($\langle \bf{v}{\bf \times} {\bf b}\rangle \equiv \xi$) can be represented by a linear combination of ${\bf \overline{B}}$ and ${\bf b}_i$ such as ${\bf B}$, ${\bf \nabla}{\bf \times}{\bf B}$, ${\bf b}_i$, and ${\bf \nabla}{\bf \times}{\bf b}_i$. Thus, we assume the basic structure of EMF is
\begin{eqnarray}
\xi=\xi_1+\xi_2=\alpha_1 {\bf B}-\beta_1 {\bf \nabla} {\bf \times}{\bf B}+\alpha_2 {\bf b}_i-\beta_2 {\bf \nabla} {\bf \times}{\bf b}_i.
\label{modified EMF alpha beta}
\end{eqnarray}
For $\xi$, we calculate $\partial /\partial t\langle {\bf u} \times {\bf b}\rangle=\langle \partial {\bf u}/\partial t \times {\bf b}\rangle+\langle {\bf u} \times \partial {\bf b}/\partial t\rangle$ to use the known momentum and magnetic induction equation.\\

\noindent After the simulation begins with the large ${\bf b}_i$(0) or $E_M(0)$, $E_M(0)$ decreases very quickly as ${\bf u}$ grows. In a few time unit ($t\sim 5$) ${\bf u}$ gets almost saturated, but ${\bf b}$ is still growing. Thus, we start the calculation using $\langle {\bf u} \times \partial {\bf b}/\partial t\rangle$($\sim \xi_k$).
From Eq.(\ref{MagIndSmall3}),
\begin{eqnarray}
{\bf b}(t')
&=&\int^{t'} \bigg({\bf \nabla}{\bf \times}({\bf u}{\bf \times}\overline{\bf B})+{\bf \nabla}{\bf \times}({\bf u}{\bf \times}{\bf b}_i)\bigg)\,d\tau.
\label{MagIndSmall4}
\end{eqnarray}
Since the basic structures of $\xi_{k,1}(B)$ and $\xi_{k,2}(b_i)$ are the same, we calculate $\xi_{k,2}$ and then change the variables to get $\xi_{k,1}$. $\xi_{k,2}(b_i)$ is
\begin{eqnarray}
\xi_{k,2}&=&\int_{-\infty}^t \overline{u({\bf x}, t){\bf \times}\bigg[{\bf \nabla}{\bf \times}
\big({\bf u}({\bf x}, t'){\bf \times}{\bf b}_i({\bf x}, t')\big)\bigg]}\,dt'.
\label{EMF soulution2}
\end{eqnarray}
Using Eq.(\ref{modified EMF alpha beta}), we get $\xi_{k,2x}$.
\begin{eqnarray}
\xi_{k,2x}&=&\bigg(u_y\frac{\partial u'_z}{\partial x}-u_z\frac{\partial u'_y}{\partial x}\bigg)b_{ix}-u_yu'_y\frac{\partial b_{iz}}{\partial y}
+u_zu'_z\frac{\partial b_{iy}}{\partial z}\nonumber\\
&\equiv& \alpha_{k,2x}b_{ix}-\beta_{k,2x}({\bf \nabla}{\bf \times} b_i)_x
\label{simplified EMF x components}
\end{eqnarray}
$\xi_{k,2y}$ has the same structure but the variables rotate: $x$$\rightarrow$$y$, $y$$\rightarrow$$z$, $z$$\rightarrow$$x$. And for $\xi_{k,2z}$, $x$$\rightarrow$$z$, $y$$\rightarrow$$x$, $z$$\rightarrow$$y$. Since we assume the system is isotropic, the coefficients of $b_{ix}$, $b_{iy}$, $b_{iz}$ are the same.
\begin{eqnarray}
u_y\frac{\partial u'_z}{\partial x}-u_z\frac{\partial u'_y}{\partial x}=u_z\frac{\partial u'_x}{\partial y}-u_x\frac{\partial u'_z}{\partial y}=u_x\frac{\partial u'_y}{\partial z}-u_y\frac{\partial u'_x}{\partial z}\nonumber\\
\Rightarrow\frac{1}{3}\bigg(u_y\frac{\partial u'_z}{\partial x}-u_z\frac{\partial u'_y}{\partial x}+u_z\frac{\partial u'_x}{\partial y}-u_x\frac{\partial u'_z}{\partial y}+u_x\frac{\partial u'_y}{\partial z}-u_y\frac{\partial u'_x}{\partial z}\bigg).
\label{simplified EMF alpha coefficient 1}
\end{eqnarray}
Then, $\alpha_{k,2}$ is
\begin{eqnarray}
\alpha_{k,2}=-\frac{1}{3}\int ^t_{-\infty}\overline{{\bf u}(x,t)\cdot {\bf \nabla} {\bf \times} {\bf u}(x,t')}\, dt'.
\label{alpha coefficient 2}
\end{eqnarray}
Similarly,
\begin{eqnarray}
u_xu'_x=u_yu'_y=u_zu'_z\Rightarrow \frac{1}{3}\big(u_xu'_x+u_yu'_y+u_zu'_z\big).
\label{simplified EMF beta coefficient 1}
\end{eqnarray}
\begin{eqnarray}
\beta_{k,2}=\frac{1}{3}\int ^t_{-\infty}\overline{{\bf u}(x,t)\cdot {\bf u}(x,t')}\, dt'
\label{alpha coefficient 2}
\end{eqnarray}

\noindent The coefficients $\alpha_{k,1}$ and $\beta_{k,1}$ are the same as mentioned.\\

\noindent While ${\bf b}$ field is even larger than growing velocity field, or stationary ${\bf b}$ ($2\leq k\leq k_{diss}$) is large enough to affect the plasma motion, we calculate $\langle \partial {\bf u}/\partial t \times {\bf b}\rangle$.

\noindent We assume that dissipation effect is ignorably small and Lorentz force is a dominant term in the momentum equation. Then,
\begin{eqnarray}
\frac{\partial {\bf U}}{\partial t}&\sim&{\bf J}{\bf \times} {\bf B}=({\bf \overline{J}}+{\bf j}_i+{\bf j}){\bf \times} ({\bf \overline{B}}+{\bf b}_i+{\bf b}).
\label{alpha coefficient by magnetic field 1}
\end{eqnarray}
Small scale momentum equation is,
\begin{eqnarray}
\frac{\partial {\bf u}}{\partial t}&\sim&{\bf \overline{B}}\cdot{\bf \nabla}{\bf b}+{\bf b}_i\cdot{\bf \nabla}{\bf b}.
\label{alpha coefficient coefficient by magnetic field 3}
\end{eqnarray}
Here we assume the averages of ${\bf b}_i$ and $\overline{{\bf B}}$ are not zero, and their spatial changes are ignorably small within the small scale eddy turnover time. Then $EMF$($\xi_M=\xi_{M,1}(\overline{B})+\xi_{M,2}(b_i)$) is,
\begin{eqnarray}
{\bf u}{\bf \times} {\bf b}&=&\int_{-\infty}^t\bigg[{\bf \overline{B}}(x,t)\cdot{\bf \nabla}{\bf b}(x,t')\,dt'\bigg] \times {\bf b}(x,t)\nonumber\\
&&+\int_{-\infty}^t\bigg[{\bf b}_i(x,t)\cdot{\bf \nabla}{\bf b}(x,t')\, dt'\bigg]\times {\bf b}(x,t)
\label{alpha coefficient coefficient by magnetic field 4}
\end{eqnarray}
The integrands of $\xi_{M,1x}$ and $\xi_{M,2x}$ are of the same structure. So if we consider $\xi_{M,1x}$,
\begin{eqnarray}
\xi_{M, 1x}\sim B_x\frac{\partial b'_y}{\partial x}b_z-B_x\frac{\partial b'_z}{\partial x}b_y.
\label{alpha coefficient coefficient by magnetic field x component}
\end{eqnarray}
$\xi_{M, 1y}$ and $\xi_{M, 1z}$ have the same results with the rotation of variables mentioned before. Also the assumption of isotropy makes the results simple.
\begin{eqnarray}
&&\frac{\partial b'_y}{\partial x}b_z-\frac{\partial b'_z}{\partial x}b_y=\frac{\partial b'_z}{\partial y}b_x-\frac{\partial b'_x}{\partial y}b_z=\frac{\partial b'_x}{\partial z}b_y-\frac{\partial b'_y}{\partial z}b_x\nonumber\\
&&\Rightarrow\frac{1}{3}\bigg(\frac{\partial b'_y}{\partial x}b_z-\frac{\partial b'_z}{\partial x}b_y+\frac{\partial b'_z}{\partial y}b_x-\frac{\partial b'_x}{\partial y}b_z+\frac{\partial b'_x}{\partial z}b_y-\frac{\partial b'_y}{\partial z}b_x\bigg)\nonumber\\
&&=\frac{1}{3}{\bf b}\cdot {\bf \nabla} {\bf \times} {\bf b'}
\label{alpha coefficient coefficient by magnetic field all components}
\end{eqnarray}
Thus, $\alpha_{M,1}$ related to ${\bf \overline{B}}$ field is
\begin{eqnarray}
\alpha_{M,1}(=\alpha_{M,2})=\frac{1}{3}\int ^t_{-\infty}\overline{{\bf b}(x,t)\cdot {\bf j}(x,t')}\, dt'.
\label{alpha coefficient due to magnetic field}
\end{eqnarray}
Finally, the complete $EMF$ is `$\xi=\xi_{k,1}+\xi_{k,2}+\xi_{M,1}+\xi_{M,2}$'.
\begin{eqnarray}
\xi&=&\frac{1}{3}\big(\langle {\bf j}\cdot {\bf b} \rangle-\langle {\bf u}\cdot \omega \rangle \big)\tau{\bf \overline{B}}-\frac{1}{3}\langle u^2 \rangle \tau {\bf \nabla}{\bf \times}{\bf \overline{B}}\nonumber\\
&&+\frac{1}{3}\big(\langle {\bf j}\cdot {\bf b} \rangle-\langle {\bf u}\cdot \omega \rangle \big)\tau{\bf b}_i(0)e^{-\eta k_f^2t}-\frac{1}{3}\langle u^2 \rangle \tau {\bf \nabla}{\bf \times}{\bf b}_i(0)e^{-\eta k_f^2t}.
\label{MagIndSmall4}
\end{eqnarray}
($\tau$ is substituted for the integration. Only the magnitude is considered.)\\ 
$E_{M,L}$ in Eq.(\ref{Em2}) is,
\begin{eqnarray}
\frac{\partial }{\partial t}E_{M,L}&=&\alpha k^2H_{M,L}-2k^2\big(\beta+\eta\big)E_{M,L}\nonumber\\
&&+\overline{{\bf B}}\cdot {\bf \nabla}\times \alpha{\bf b}_i(0)e^{-\eta\,k^2_ft}+\beta\,\overline{{\bf B}}\cdot \nabla^2{\bf b}_i(0)e^{-\eta\,k^2_ft}.
\label{Emfinal}
\end{eqnarray}
The first and third term on the right hand side are the sources of $E_{M,L}$. These two terms describe the inverse cascade of energy in small scale to $E_{M,L}$ with $\alpha$. In fact, Fourier transformed representation shows the mean correlation $\langle \overline{{\bf B}}\cdot {\bf \nabla}\times {\bf b}_i(0) \rangle$ has a nontrivial value only with $\alpha$. We use Fourier transformation `$f(x)=\int\, f({\bf k}) e^{i{\bf k}\cdot {\bf x}}\, d{\bf k}$' and `$\partial E_{M,L}/\partial t=1/2[\overline{{\bf B}}(-k)\cdot \partial \overline{{\bf B}}(k)/\partial t+\overline{{\bf B}}(k)\cdot \partial \overline{{\bf B}}(-k)/\partial t$]'. Then, $\overline{{\bf B}}\cdot {\bf \nabla}\times \alpha{\bf b}_i$ is,
\begin{eqnarray}
&\sim&\frac{1}{2}\sum_{p,\,q}\bigg[\frac{1}{3}\bigg({\bf j}(p)\cdot {\bf b}(q) - {\bf u}(p)\cdot \omega(q)  \bigg)\tau \epsilon_{i'lm} \frac{\partial}{\partial x_l} {\bf b}_{i,m}(k_f)\overline{B}_{i'}(-k)+\nonumber\\
&&\frac{1}{3}\bigg({\bf j}(-p)\cdot {\bf b}(-q) - {\bf u}(-p)\cdot \omega(-q)  \bigg)\tau \epsilon_{i'lm} \frac{\partial}{\partial x_l} {\bf b}_{i,m}(-k_f)\overline{B}_{i'}(k)\bigg]\nonumber\\
&=& \frac{1}{3} \mathfrak{Im}\bigg[\sum_{p+q+k_f=k}\bigg({\bf u}(p)\cdot \omega(q)-{\bf j}(p)\cdot {\bf b}(q)\bigg)\,\epsilon_{i'lm} k_{f,l}{\bf b}_{i,m}(k_f)\overline{B}_{i'}(-k) \bigg]\tau
\label{MagEnergyFourier1}
\end{eqnarray}
The current helicity and kinetic helicity whose wave numbers satisfy the relation $p+q=-4$ ($|p|,\,|q|\geq 2$) contribute to the growth of large scale magnetic energy. As Fig.\ref{2d} shows, the growth rate of $E_{M,L}$ or $H_{M,L}$ is the largest when $H_M$ at $k\sim 5$ is positive(right handed) and $H_M$ at $k\sim1-2$ is negative(left handed). Then negative $\langle {\bf j}(p)\cdot {\bf b}(q) \rangle$ increases the magnitude of $\alpha$ coefficient ($\sim (\langle {\bf j}(p)\cdot {\bf b}(q) \rangle$-$\langle {\bf v}(p)\cdot {\bf \omega}(q) \rangle)\tau$). The difference of $p$ and $q$ here does not exactly satisfy the criterion, but this method explains the simulation results quite well. In fact, the interaction among eddies in real turbulence is not so strictly limited as the theoretical inference predicts. On the other hand the relation of $p,\,q$ for $\alpha k^2H_{M,L}$($\langle \overline{{\bf B}}\cdot {\bf \nabla}\times \alpha\overline{{\bf B}} \rangle$) is `$p+q=0$'. The sign of $\langle {\bf j}(p)\cdot {\bf b}(q)\rangle(=p^2\langle {\bf a}(p)\cdot {\bf b}(q)\rangle)$ is always opposite to that of $\langle {\bf v}(p)\cdot {\bf \omega}(q) \rangle$.\\


\noindent For the dynamo without helicity, the above equations cannot be used; and, ${\bf b}_i$ cannot directly interact with $\overline{{\bf B}}$. Instead, we should use $\overline{{\bf B}}\cdot {\bf \nabla}\times \langle {\bf u}\times {\bf b}\rangle$. The source term ($EMF$) is (Kraichnan and Nagarajan 1967)
\begin{eqnarray}
\sim \mathfrak{Im}\bigg[k_m\sum_{k'=2}^{k_{max}}\bigg(\big<\, \overline{{\bf B}}(-k)\cdot {\bf b}(k')u_m(k-k')\big>-\big<\,b_{m}(k')\overline{{\bf B}}(-k)\cdot {\bf u}(k-k')\big>\bigg)\,\bigg].\quad (k=1)
\label{MagEnergyFourier2}
\end{eqnarray}
This equation is more exact and general than Eq.(\ref{MagEnergyFourier1}) whether or not the field is helical. But it is rather difficult to understand its physical meaning intuitively using this result.\\

Up to now we have used the fact that the left handed magnetic helicity($\langle {\bf a}_2\cdot {\bf b}_2\rangle$$<0$) is generated when the system is driven by the right handed kinetic helicity($\langle {\bf u}_2\cdot {\bf \omega}_2\rangle>0$) without enough consideration. Mathematically the growth of larger scale magnetic field(${\bf B}_1$) or helicity(${\bf H}_1$) is described by a differential equation like Eq.(\ref{magnetic induction 2}) or Eq.(\ref{Hm2}). However, since the differential equation in itself cannot describe the change of sign of variables, more fundamental and physical approach is necessary. In case of $\alpha$$\Omega$ dynamo, there was a trial to explain the handedness of twist and writhe in corona ejection using the concept of magnetic helicity conservation (Blackman \& Field 2003). But, even when the effect of differential rotation cannot be expected ($\alpha^2$ dynamo), the sign of generated $H_M$ and injected $\langle {\bf u}\cdot {\bf \omega}\rangle$ is opposite.\\

\noindent We assume the magnetic field $B_{1}\hat{y}$(Fig.\ref{4a}, Krause \& R\"adler 1980) interacts with right handed helical kinetic plasma motion. The velocity can be divided into toroidal component $u_{2,t}\hat{\phi}$ and poloidal component $u_{2,p}\hat{z}$. The interaction of this toroidal motion with ${\bf B}_{1}$ produces ${\bf j}_{2}$$\sim{\bf u}_{2,t}{\bf \times} {\bf B_{1}}$. The induced current density ${\bf j}_{2,f}$ in the front is toward positive $\hat{z}$ direction, but the rear current density ${\bf j}_{2,r}$ is along with the negative $\hat{z}$. These two current densities become the sources of magnetic field $-{b}_{2}$$\hat{x}$(${\bf j}\sim {\bf \nabla} {\bf \times} {\bf b}$). Again this induced magnetic field interacts with the poloidal kinetic velocity $u_{2,p}\hat{z}$ and generates -$J_{in}\hat{y}$($\sim{\bf u}_{2,p}{\bf \times} {\bf b_{2}}$). Finally this ${\bf J}_{in}$ produces ${\bf B}_{in}$, which forms a circle from the magnetic field ${\bf B}_1$(upper picture in Fig.\ref{4b}). If we go one step further from here, we see ${\bf B}_{in}$ can be considered as a new toroidal magnetic field ${\bf B}_{tor}$, and ${\bf B}_1$ as ${\bf B}_{pol}$. This new helical magnetic field structure has the left handed polarity, i.e., $\langle {\bf a}\cdot {\bf b}\rangle$$<$0 (lower plot of Fig.\ref{4b}). ${\bf B}_{tor}$ interacts with the positive $\langle {\bf u}_2\cdot {\bf \omega}_2\rangle$ and induces the current density ${\bf J}$ which is antiparallel to ${\bf B}_{tor}$. Then ${\bf B}_{pol}$ is reinforced by this ${\bf J}$, which is the typical $\alpha^2$($B_{pol}\leftrightarrow B_{tor}$) dynamo with the external forcing source.\\

\section{Conclusion}
We have seen how the initially given magnetic energy $E_M$(0) and $H_M$(0) generate the additional terms ($\sim{\bf b}_i$) in $EMF$ and affect the growth rate of large scale ${\bf B}$ field. Nontrivial interaction between ${\bf b}_i$ and $\overline{{\bf B}}$ occurs with $\alpha$ coefficient, which leads to the increase of the growth rate of $E_{M,L}$. Simulation results show the growth rate of large scale magnetic field is proportional to $E_M$(0) and positive $H_M$(0). $E_M$(0) was found to be a more important factor than $H_M$(0) in MHD dynamo. As ${\bf b}_i (\sim e^{-\eta k^2_ft})$ implies, the saturated value is independent of these initial conditions. We have also seen the physical role and complimentary relation between helical and nonhelical magnetic fields. The helical kinetic and magnetic field are related to the inverse cascade of the magnetic energy. The nonhelical magnetic fields can generate the helical magnetic fields and constrain the plasma motion through Lorentz force. Not much about nonhelical magnetic field has been known yet. In this paper, we assumed a homogeneous and isotropic system for simplicity. However, if there is a mean or large scale magnetic field in the system, the kinetic and magnetic field is not isotropic anymore, which leads to the modification of $\alpha^2$ dynamo model. We will leave this topic for the future work.\\




\section{Acknowledgement}
\noindent KWP acknowledges support from the National Research Foundation of Korea through grant 2007-0093860. KWP appreciates the comments from Dr. Dongsu Ryu at UNIST.

\section{References}
\noindent Biskamp, D. 2003, Magnetohydrodynamic Turbulence (Cambridge press)\\
Blackman, E. G. \& Brandenburg, A. 2003, ApJL, 584, L99\\
Blackman, E. G. \& Field, G. B. 2002, PRL, 89, 265007\\
Brandenburg, A. 2001, ApJ, 550, 824\\
Brown, M. R. \& Canfield, R. C. \& Pevtsov, A. A. 1999, Magnetic Helicity in Space and Laboratory Plasmas (American Geophysical Union)\\
Frisch, U., Pouquet, A., Leorat, J., \& Mazure, A. 1975, J. Fluid Mech, 68, 769\\
Goldreich, P., \& Sridhar, S. 1995, ApJ, 438, 763
Iroshnikov, P. S. 1964, Sovast, 7, 566\\
Kraichnan, R. H. 1965, Phys. Fluids, 8, 1385\\
Kraichnan, R. H., \& Nagarajan, S. 1967, Phys. Fluids, 10, 859\\
Krause, F. \& R\"adler, K. H. 1980, Mean-field magnetohydrodynamics and dynamo theory (Pergamon)\\
Lesieur, M. 1987, Turbulence in Fluids: Stochastic and Numerical Modeling (2nd ed.; Springer)\\
Maron, J. \& Blackman, E. G. 2002, ApJL, 566, L41\\
Moffatt, H. K. 1978, Magnetic Field Generation in Electrically Conducting Fluids (Cambridge University Press)\\
Park, K. 2013, MNRAS, 434, 2020\\
Park, K. \& Blackman, E. G. 2012, MNRAS, 419, 913\\
Park, K. \& Blackman, E. G. 2012, MNRAS, 423, 2120\\
Park, K., Blackman, E. G., \& Subramanian 2013, PRE, 87, 053110\\
Pouquet, A., Frisch, U., \& Leorat, J. 1976, J. Fluid Mech., 77, 321\\
Robertson, H. P. 1940, Proc. Cambridge Philos. Soc., 36, 209\\  
Yoshizawa, A. 2011, Hydrodynamic and Magnetohydrodynamic Turbulent Flows: Modeling and Statistical Theory (Fluid Mechanics and Its Applications) (Springer)\\


\end{document}